\documentclass[a4paper,twocolumn,fleqn]{article}

\usepackage{txfonts}               
 \usepackage{cite}                  
\usepackage[dvipdfm]{graphicx}     
\usepackage{pfr}                   

\begin{document}

\title{Hydrogen Adsorption of Back Side of Graphene}


\author{Atsushi~ITO\sup{1}, Arimichi~TAKAYAMA\sup{2} and Hiroaki~NAKAMURA\sup{2}}

\affiliation{\sup{1}Department of Physics, Graduate School of Science, Nagoya University, Furo--cho, Chikusa--ku, Nagoya 464--8602, Japan. \\
  \sup{2}National Institute for Fusion Science, Oroshi--cho 322--6, Toki 509--5292, Japan.}

\date{(Received 15 October 2007 / Accepted ** ** 2007)}

\email{ito.atsushi@nifs.ac.jp, nakamura.hiroaki@nifs.ac.jp, takayama@nifs.ac.jp}

\begin{abstract}
We studied the interaction between a single hydrogen atom and a single graphene using classical molecular dynamics simulation with modified Brenner REBO potential.
Three interactions, which are adsorption, reflection, penetration, were observed.
Overhang structure appears and creates an adsorption site on the backside of the graphene.
It is considered that backside adsorption occurs under the two conditions that an incident hydrogen atom should have incident energy which is larger than the potential barrier of a hexagonal hole of the graphene and that after the hydrogen atom passes through the graphene, it does not keep its kinetic energy to be trapped by the adsorption site.
The conditions explained that as the incident energy increased, the incident point of the backside adsorption shifted to the periphery of a hexagonal hole of the graphene in the simulation.
Moreover, when a hexagonal hole of the graphene was expanded by the hydrogen atom incidence to the periphery of the hexagonal hole, its potential barrier was reduced.\\
\end{abstract}


\DOI{****/pfr.****}

\maketitle  

\section{Introduction}

In the context of research into nuclear fusion, the plasma surface interaction (PSI) problem has been studied \cite{Nakano,Roth,Roth2,Mech,LHD}.
A portion of the plasma confined in an experimental device falls onto a divertor wall, which is shielded by graphite or carbon fiber composite tiles.
The incident hydrogen plasma erodes these carbon tiles in a process called chemical sputtering.
The erosion produces hydrocarbon molecules, such as $\textrm{CH}_{\mathrm{x}}$ and $\textrm{C}_2\textrm{H}_{\mathrm{x}}$, which affect the plasma confinement.

To solve the PSI problem, the mechanism of the graphite erosion has been researched using molecular dynamics (MD) simulation \cite{Salonen,Salonen2,Alman,Marian}.
Previously, we investigated the PSI of graphite surface which consists of 8 graphene sheets using the modified Brenner reactive empirical bond order (REBO) potential \cite{Nakamura}.
Subsequently, the isotope dependence of incident hydrogen atoms is investigated \cite{Ito_mghdt}.
These simulations achieve steady state of the graphite erosion, the incident energy linear dependence of total carbon yield accords with experimental results \cite{Mech2, Garcia}.

The MD simulation of graphite surface showed that if incident energy was 5 eV, almost of all incident hydrogen atoms were adsorbed by the graphite surface, while if the incident energy was 15 eV, most incident hydrogen atoms were reflected.
This adsorption and reflection can be explained by the MD simulation of the elemental processes which is the chemical reaction between a single hydrogen atom and a single graphene \cite{Ito_ICNSP, Ito_gh1, Nakamura_PET}.
This MD simulation indicates also that in a certain incident energy, the hydrogen atom can be adsorbed to the backside of the graphene.
On the other hand, the other MD simulation for graphene erosion due to hydrogen atom gas implied that para--overhang configuration which has hydrogen atoms onto both side of the graphene is created and its C--C bond is easy broken \cite{Ito_gh2}.
Therefore, we consider that if the incident energy which brings about the backside adsorption is selected, the MD simulation of graphite surface shows a different dynamics.

Here, we attach weight to not only the macroscopic simulation of graphite surface but also understanding a elemental mechanism and dynamics. 
We, therefore, investigate the backside adsorption of the interaction between a single hydrogen atom and a single graphene in the present paper.
We describe the simulation model and method in \S \ref{ss:SimMethod}.
In \S \ref{ss:Result}, we present and discuss the simulation results.
This paper concludes with a \S \ref{ss:Summary}.

\section{Simulation Method}\label{ss:SimMethod}

A graphene \cite{Boehm} consists of 160 carbon atoms measuring 2.13 nm $\times$ 1.97 nm, and is placed at the center of simulation box parallel to $x$--$y$ plane.
The simulation box in the $x$-- and $y$--directions measures 2.13 nm $\times$ 1.97 nm with periodic boundary condition.
In the $z$--direction, we do not prepare the boundary of the simulation box and the initial coordinate of the center of mass of the graphene is $z = 0$~\AA.
The initial graphene temperature is set to 0 K.
The carbon atoms of the graphene are relaxed initially and compose completely flat graphene structure.
A hydrogen atom is injected from $z = 4$~\AA~parallel to the $x$--$y$ plane, that is, normal incidence to the graphene surface.
Incident position in the $x$-- and $y$-- directions are determined under a uniformed distribution function.
The incident energy $E_\mathrm{I}$ decides an initial momentum vector (0, 0, $p_0$) as follows:
\begin{eqnarray}
	p_0 = \sqrt{2 m E_{\mathrm{I}}}, \label{eq:inip}
\end{eqnarray}
where and $m$ is the mass of the hydrogen atom.

We performed our MD simulation under \textit{NVE} conditions, where the number of atoms, volume, and total energy are conserved.
The simulation time was developed using second order symplectic integration \cite{Suzuki}.
The chemical interaction was represented by the modified Brenner REBO potential \cite{Ito_gh1, Brenner}:
\begin{eqnarray}
	U \equiv \sum_{i,j>i} \Bigg[V_{[ij]}^\mathrm{R}( r_{ij} )
		 - \bar{b}_{ij}(\{r\},\{\theta^\mathrm{B}\},\{\theta^\mathrm{DH}\}) V_{[ij]}^\mathrm{A}(r_{ij}) \Bigg],
	\label{eq:model_rebo}
\end{eqnarray}
where $r_{ij}$ is the distance between the $i$--th and $j$--th atoms.
The functions $V_{[ij]}^{\mathrm{R}}$ and $V_{[ij]}^{\mathrm{A}}$ represent repulsion and attraction, respectively.
The function $\bar{b}_{ij}$ generates multi--body force.
To conserve the accuracy of the calculation, the time step was $5\times10^{-18} \mathrm{~s}$.

\begin{figure}[t]
\centering
		\resizebox{\linewidth}{!}{\includegraphics{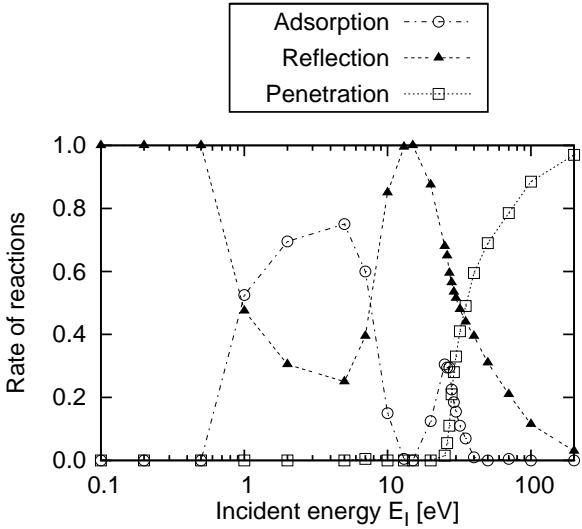}}
\caption{The incident energy dependence of the rates of adsorption, reflection and penetration where the graphene temperature is 0 K.
	  Dash--dotted line with	open circle, long--dashed line with filled triangle, and short--dashed line with square denote the rates of adsorption, reflection and penetration, respectively.}
\label{fig:XX1}
\end{figure}

We note the condition to judge the type of reaction, which corresponds to the condition to finish the simulation.
While the incident hydrogen atom interacts with one carbon atom, we count a trapped time.
When the hydrogen atom leaves the carbon atom or starts interaction with the other carbon atom, the trapped time is cleared.
When the trapped time reaches 0.1 ps (20000 time steps), the simulation is finished and this reaction is regarded as adsorption.
Moreover, if the relative position from the nearest carbon atom to the hydrogen atom in the $z$--coordinate is positive, the reaction is front adsorption, while if the relative position in the $z$--coordinate is negative, the reaction is backside adsorption.
When the hydrogen atom leaves the nearest carbon atom, we begin counting a escaping time.
If the hydrogen atom interacts a carbon atom again, we clear the escaping time and restart counting the trapped time.
When the escaping time reaches 0.05 ps (10000 time steps), the simulation is finished.
And, if the escaping hydrogen atom has positive momentum in the $z$--coordinate, this reaction is reflection, while it has negative momentum, the reaction is penetration.

We repeated the above simulation 200 times for each incident energy, where the incident position in the $x$-- and $y$-- directions are every time changed randomly under the uniformed distribution function.
Since we counted the types of the reactions, we obtained the rates of the reactions.
Of course, the sum of the rates of the adsorption, reflection and penetration is always 1.

\section{Results and Discussion}\label{ss:Result}

\begin{figure}[t]
\centering
		\resizebox{\linewidth}{!}{\includegraphics{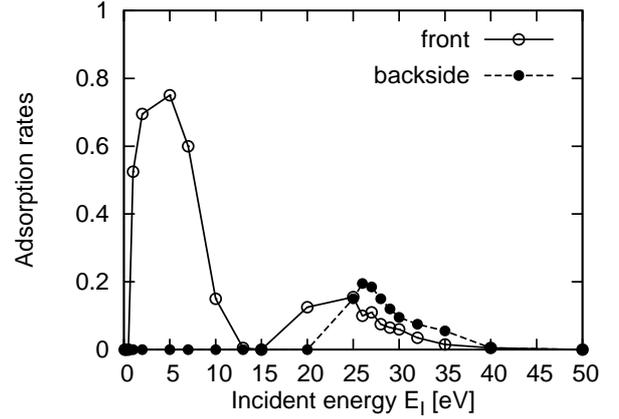}}
\caption{The incident energy dependence of the rates of the front and backside adsorption where the graphene temperature is 0 K.}
\label{fig:XX2}
\end{figure}

The simulations were executed for incident energy of 0.1 eV to 200 eV.
The three interactions, which are adsorption, reflection, penetration, were observed.
Figure \ref{fig:XX1} shows that the rates of the three interactions depend on the incident energy.
Especially, the rate of adsorption has two peaks.
Next, Fig. \ref{fig:XX2} shows the rates of the front and backside adsorption.
Although the front adsorption occurs around the both adsorption peaks, the backside adsorption occurs around the high incident energy adsorption peak only.
In the process of front adsorption, the nearest carbon atom which is connected to the incident hydrogen atom by a covalent bond is pulled out from flat surface of the graphene.
This is called a overhang structure and creates an adsorption site for a hydrogen atom \cite{Ito_ICNSP,Ito_gh1}.
In the backside adsorption also, the overhang structure appears.
That is, an adsorption site is created in the backside of graphene after the incident hydrogen atom goes through the graphene.

\begin{figure}[tb]
\centering
		\resizebox{\linewidth}{!}{\includegraphics{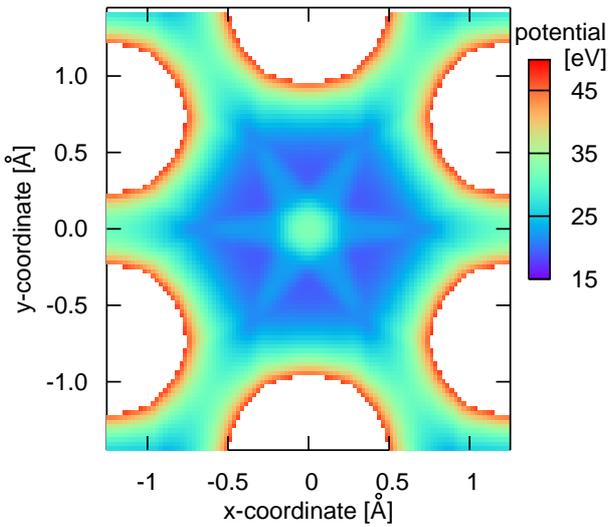}}
\caption{The potential barrier of a hexagonal hole of the graphene.
The mapped value is potential energy when the hydrogen atom is located in the plane of $z = 0$~\AA.}
\label{fig:XX5}
\end{figure}

\begin{figure}[tb]
\centering
\begin{tabular}{cc}
		\resizebox{0.5\linewidth}{!}{\includegraphics{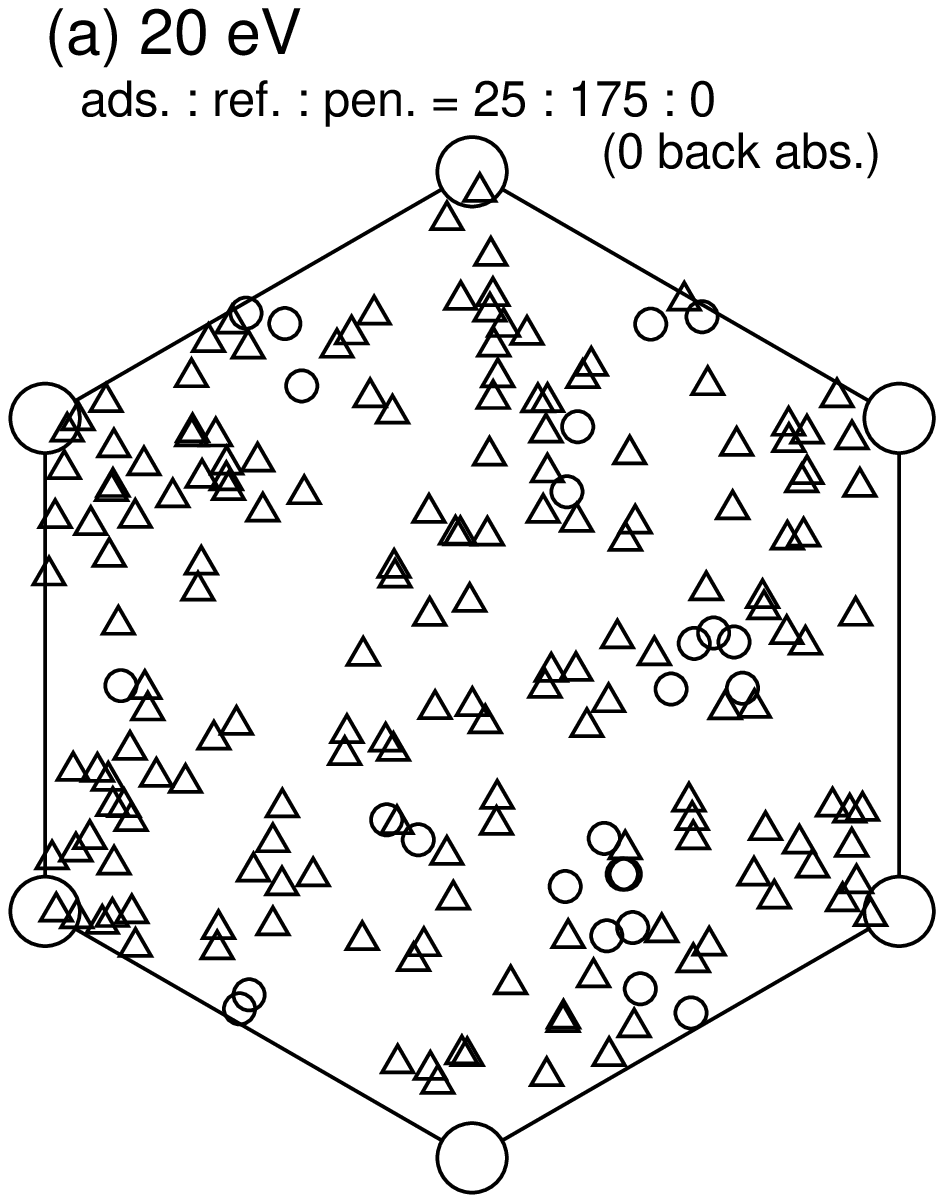}} &
		\resizebox{0.5\linewidth}{!}{\includegraphics{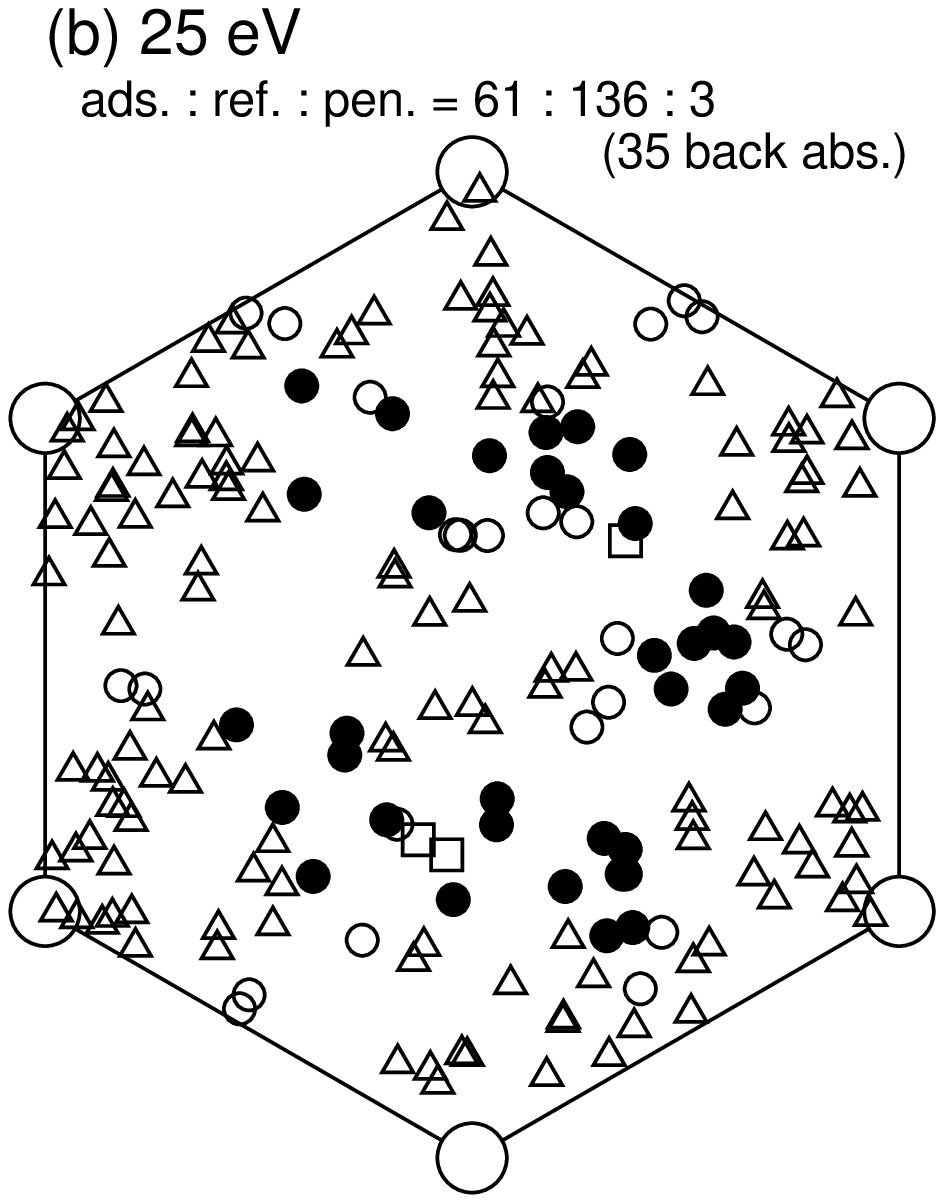}} \\
		\resizebox{0.5\linewidth}{!}{\includegraphics{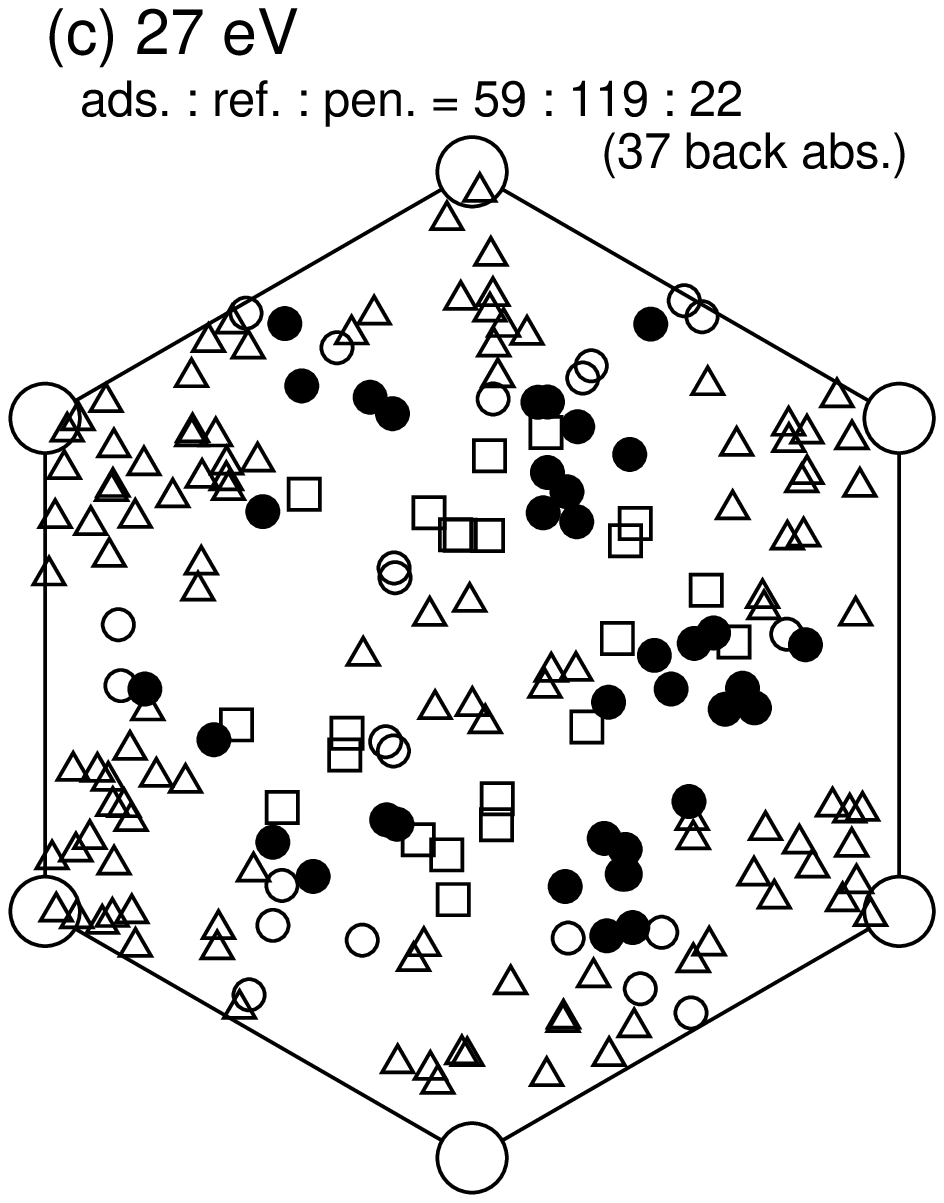}} &
		\resizebox{0.5\linewidth}{!}{\includegraphics{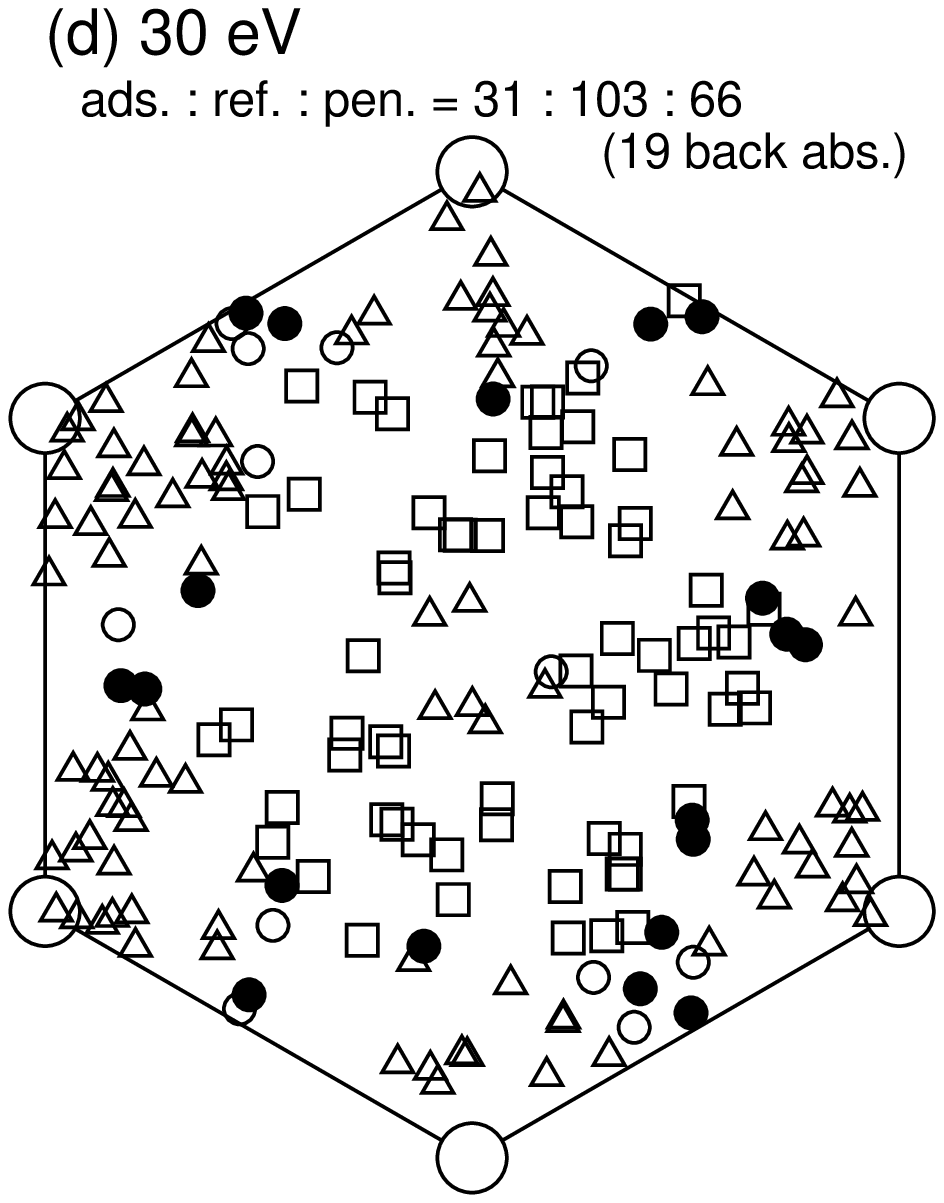}} \\
	\end{tabular}
\caption{Incident points v.s. the types of its interactions.
Circles indicate incident points which bring about hydrogen atom adsorption where the opaque circles correspond to backside adsorption.
Triangles and squares are reflection and penetration, respectively.
Big six circles and six line represent six carbon atoms and six covalent bonds composing a hexagonal hole of the graphene.}
\label{fig:XX4}
\end{figure}

The incident energy dependence of the rates of interactions Fig. \ref{fig:XX1} is understood using energetics, which relates $\pi$--electron on the graphene surface, structure transformation to the overhang structure and the potential barrier of a hexagonal hole of the graphene \cite{Ito_gh1}.
In the present paper, we discuss the backside adsorption.
We consider that the backside adsorption needs the following two conditions.
One is that to go through the graphene, the incident hydrogen atom should have the incident energy which is larger than the potential barrier of a hexagonal hole of the graphene.
The other is that after the hydrogen atom passes through the graphene, it is trapped by the adsorption site due to the overhang structure.
Namely, to occur the backside adsorption, the kinetic energy of the hydrogen atom is diffused to carbon atoms of the graphene and must not remain.
Concerning the first condition, the potential barrier of a hexagonal hole of the graphene, where the hydrogen atom is located at $z = 0$~\AA, is shown by Fig. \ref{fig:XX5}.
The center of this figure corresponds to the center of a hexagonal hole of the graphene.
The six white regions indicate higher potential barrier of more than 50 eV because carbon atoms exist there.
The point of minimum potential barrier is not the center of a hexagonal hole of graphene and appears between the locations of carbon atoms and the center of a hexagonal hole of the graphene.
Figure \ref{fig:XX4} shows incident points and the types of their interactions.
It is understood and consistent with the potential barrier of a hexagonal hole of the graphene Fig. \ref{fig:XX5} that the backside adsorption occurs around the points of minimum potential barrier and does not occur around the center of a hexagonal hole of the graphene.
As the incident energy increases, the incident points for the backside adsorption shift from the points of minimum potential barrier to the periphery of a hexagonal hole of the graphene which is over the covalent bonds between carbon atoms.
This phenomenon is explained as follows.
In the incident energy of 20 eV, because all incidences become the front adsorption or reflection, the hydrogen atom does not have enough incident energy to go over the potential barrier of a hexagonal hole of the graphene.
In the incident energy of 25 eV, because the incident energy is higher than but close to the minimum potential barrier, only the hydrogen atoms which are injected around the points of the minimum potential barrier go through the hexagonal hole of the graphene and are trapped by adsorption site of the backside of the graphene.
When the incident energy increases (27 and 30 eV), the region in which the potential barrier of a hexagonal hole of the graphene is lower than the incident energy extends.
Therefore, the backside adsorption occurs in the periphery of a hexagonal hole of the graphene.
However, the backside adsorption around the potential minimum point changes into penetration because the hydrogen atom leaves the kinetic energy enough to escape out of the adsorption site after it passed through a hexagonal hole of the graphene.
More important thing is that while the backside adsorption happens in the periphery of a hexagonal hole of the graphene, it does not happen in the center of a hole of the graphene in spite of same potential barrier.
It is demonstrated by the trajectory of atoms in the MD simulation that when the hydrogen atom is injected into the periphery of a hexagonal hole of the graphene, near two carbon atoms move and the hexagonal hole of the graphene is expanded.
The expansion of the hexagonal hole of the graphene seems to reduce the potential barrier of the hexagonal hole of the graphene because the distance between hydrogen atom and the near carbon atoms become larger.
On the other hand, if the hydrogen atom is injected to the center of a hexagonal hole of the graphene, it must push six carbon atoms to expand the hexagonal hole of the graphene.
It is considered that at the incident energy of 30 eV, even if the hydrogen atom can move two carbon atoms, it cannot move six carbon atoms because of their larger mass.
Consequently, the backside adsorption and penetration hardly occur in the center of a hexagonal hole of the graphene.

%

\section{Summary}\label{ss:Summary}

We study the interaction between a single hydrogen atom and a single graphene using classical molecular dynamics simulation with modified Brenner REBO potential.
The three interactions, which are adsorption, reflection, penetration, were observed.
Especially, in the present paper, we discuss the backside adsorption.
As well as the front adsorption, the overhang structure appears in the backside of the graphene and creates the adsorption site.
It is considered that the backside adsorption occurs under the two conditions that the incident hydrogen atoms should have the incident energy which is larger than the potential barrier of a hexagonal hole of the graphene and that after the hydrogen atom passes through the graphene, it does not keep its kinetic energy to be trapped by the adsorption site.
The hydrogen atom in the incident energy of 25 eV brings about the backside adsorption only around the points of the minimum potential barrier, which is not the center of a hexagonal hole of the graphene and is the region between the center of a hexagonal hole and the locations of carbon atoms.
When the incident energy increases, the incident points for the backside adsorption shift to the periphery of a hexagonal hole of the graphene.
In addition, the backside adsorption around the potential minimum point change into penetration because the hydrogen atom leaves enough kinetic energy to escape from the adsorption site after it passed through a hexagonal hole of the graphene.
Moreover, when a hexagonal hole of the graphene is expanded by the hydrogen atom incidence to the periphery of the hexagonal hole, its potential barrier is reduced.
However, even if the hydrogen atom can move two carbon atoms in the incidence to the periphery of the hexagonal hole, it cannot move six carbon atoms in the incidence to the center of the hexagonal hole.
Therefore, the backside adsorption and penetration hardly occur in the center of a hexagonal hole of the graphene.



\section*{References}

\begin{thebibliography}{9}
	\bibitem{Nakano} T. Nakano, H. Kubo, S. Higashijima, N. Asakura, H. Takenaga, T. Sugie, K. Itami, Nucl. Fusion \textbf{42}, 689 (2004).
	\bibitem{Roth} J. Roth, J. Nucl. Mater. \textbf{266--269}, 51 (1999).
	\bibitem{Roth2} J. Roth, R. Preuss, W. Bohmeyer, S. Brezinsek, A. Cambe, E. Casarotto, R. Doerner, E. Gauthier, G. Federici, S. Higashijima, J. Hogan, A. Kallenbach, A. Kirschner, H. Kubo, J. M. Layet, T. Nakano, V. Philipps, A. Pospieszczyk, R. Pugno, R. Ruggieri, B. Schweer,  G. Sergienko, M. Stamp, Nucl. Fusion \textbf{44}, L21 (2004).
	\bibitem{Mech} B. V. Mech, A. A. Haasz, J. W. Davis, J. Nucl. Mater. \textbf{241--243}, 1147 (1997).
	\bibitem{LHD} A. Sagara, S. Masuzaki, T. Morisaki, S. Morita, H. Funaba, M. Goto, Y. Takamura, K. Nishimura, N. Noda, M. Shoji, H. Suzuki, A. Takayama, A. Komori, N. Ohyabu, O. Motojima, K. Morita, K. Ohya, J. P. Sharpe, LHD experimental group, J. Nucl. Mater. \textbf{313--316}, 1 (2003).
	\bibitem{Salonen} E. Salonen, K. Nordlund, J. Tarus, T. Ahlgren, J. Keinonen, C. H. Wu, Phys. Rev. \textbf{B 60}, R14005 (1999).
	\bibitem{Salonen2} E. Salonen, K. Nordlund, J. Keinonen, C. H. Wu, Phys. Rev. \textbf{B 63}, 195415 (2001).
	\bibitem{Alman} D. A. Alman, D. N. Ruzic, J. Nucl. Mater. \textbf{313--316}, 182 (2003).
	\bibitem{Marian} J. Marian, L. A. Zepeda--Ruiz, G. H. Gilmer, E. M. Bringaand, T. Rognlien, Phys. Scr. \textbf{T 124}, 65 (2006).
	\bibitem{Nakamura} H. Nakamura, A. Ito, Mol. Sim. \textbf{33}, 121 (2007).
	\bibitem{Ito_mghdt} A.Ito and H. Nakamura, to be published in Thin Solid Films.
	\bibitem{Mech2} B. V. Mech, A. A. Haasz and J. W. Davis, J. Nucl. Mater. \textbf{255}, 153 (1998).
	\bibitem{Garcia} C. Garcia and J. Roth, J. Nucl. Mater. \textbf{196--198}, 573 (1992).
	\bibitem{Ito_ICNSP} A. Ito, H. Nakamura, J. Plasma Phys. \textbf{72}, 805 (2006).
	\bibitem{Ito_gh1} A. Ito, H. Nakamura, A. Takayama, submitted, (preprint:NIFS-872).
	\bibitem{Nakamura_PET} H. Nakamura, A. Takayama, A. Ito, submitted, (preprint:NIFS-875).
	\bibitem{Ito_gh2} A. Ito, H. Nakamura, submitted.
	\bibitem{Boehm} H. --P. Boehm, R. Setton, E. Syumpp, Pure \& Appl. Chem. \textbf{66}, 1893 (1994).
	\bibitem{Suzuki} M. Suzuki, J. Math. Phys. \textbf{26}, 601 (1985).
	\bibitem{Brenner} D. W. Brenner, O. A. Shenderova, J. A. Harrison, S. J. Stuart, B. Ni, S. B. Sinnott, J. Phys.: Condens. Matter \textbf{14}, 783 (2002).
	
\end{thebibliography}
\end{document}